\documentclass[aps,preprint]{revtex4}
\usepackage{amsmath}
\usepackage[english]{babel}
\usepackage[latin1]{inputenc}
\usepackage{amssymb}
\usepackage{graphicx}
\usepackage{natbib}
\usepackage{epsfig}
\usepackage{bm}
\usepackage{dcolumn}

\def\tra{{\rm Tr}\,}
\def\trmin{{\rm tr}}
\def\mfa{{\mbox{\tiny MFA}}}

\begin{document}

\vspace*{2cm}

\title{\sc\Large{Neutral meson properties under an external magnetic field in nonlocal chiral quark models}}

\author{D. G\'omez Dumm$^{a,b}$, M.F. Izzo Villafa\~ne$^{a,b}$, N.N.\ Scoccola$^{b,c,d}$}

\affiliation{$^{a}$ IFLP, CONICET $-$ Dpto.\ de F\'{\i}sica, Fac.\ de Cs.\ Exactas,
Universidad Nacional de La Plata, C.C. 67, (1900) La Plata, Argentina}
\affiliation{$^{b}$ CONICET, Rivadavia 1917, (1033) Buenos Aires, Argentina}
\affiliation{$^{c}$ Physics Department, Comisi\'{o}n Nacional de Energ\'{\i}a At\'{o}mica, }
\affiliation{Av.\ Libertador 8250, (1429) Buenos Aires, Argentina}
\affiliation{$^{d}$ Universidad Favaloro, Sol{\'{\i}}s 453, (1078) Buenos Aires, Argentina}

\begin{abstract}
We study the behavior of neutral meson properties in the presence of a
static uniform external magnetic field in the context of nonlocal chiral
quark models. The formalism is worked out introducing Ritus transforms of
Dirac fields, which allow to obtain closed analytical expressions for
$\pi^0$ and $\sigma$ meson masses and for the $\pi^0$ decay constant.
Numerical results for these observables are quoted for various
parameterizations. In particular, the behavior of the $\pi^0$ meson mass
with the magnetic field is found to be in good agreement with lattice QCD
results. It is also seen that the Goldberger-Treiman and
Gell-Mann-Oakes-Renner chiral relations remain valid within these models in
the presence of the external magnetic field.
\end{abstract}


\maketitle

\renewcommand{\thefootnote}{\arabic{footnote}}
\setcounter{footnote}{0}

\section{Introduction}

The study of the behavior of strongly interacting matter under intense
external magnetic fields has gained increasing interest in the last few
years, especially due to its applications to the analysis of relativistic
heavy ion collisions~\cite{HIC} and the description of compact objects like
magnetars~\cite{duncan}. From the theoretical point of view, addressing this
subject requires to deal with quantum chromodynamics (QCD) in
nonperturbative regimes, therefore present analyses are based either in the
predictions of effective models or in the results obtained through lattice
QCD (LQCD) calculations. In this work we focus on the effect of an intense
external magnetic field on $\pi^0$ and $\sigma$ meson properties. This issue
has been studied in the last years following various theoretical approaches
for low-energy QCD, such as Nambu-Jona-Lasinio (NJL)-like
models~\cite{Fayazbakhsh:2013cha,Fayazbakhsh:2012vr,Avancini:2015ady,Zhang:2016qrl,Avancini:2016fgq,Mao:2017wmq},
chiral perturbation theory (ChPT)~\cite{Andersen:2012zc,Agasian:2001ym} and
path integral Hamiltonians
(PIH)~\cite{Orlovsky:2013wjd,Andreichikov:2016ayj}. In addition, results for
the light meson spectrum under background magnetic fields have been recently
obtained from LQCD calculations~\cite{Bali:2015vua,Bali:2017ian}.

We will study in particular the behavior of the mass and decay constant of
the $\pi^0$ meson in the presence of a uniform static magnetic field,
within a relativistic chiral quark model in which quarks interact through a
nonlocal four-fermion coupling~\cite{Rip97}. This so-called ``nonlocal NJL
(nlNJL) model'' can be viewed as a sort of extension of the NJL model that
intends to provide a more realistic effective approach to QCD. Actually,
nonlocality arises naturally in the context of successful descriptions of
low-energy quark dynamics~\cite{Schafer:1996wv,RW94}, and it has been
shown~\cite{Noguera:2008} that nonlocal models can lead to a momentum
dependence in quark propagators that is consistent with LQCD results.
Moreover, in this framework it is possible to obtain an adequate description
of the properties of light mesons at both zero and finite
temperature~\cite{Noguera:2008,Bowler:1994ir,Schmidt:1994di,Golli:1998rf,
General:2000zx,Scarpettini:2003fj,GomezDumm:2006vz,
Contrera:2007wu,Hell:2008cc,Dumm:2010hh,Carlomagno:2013ona}.

The basic theoretical formalism required for the study of nlNJL models in
the presence of a uniform static magnetic field $B$ has been introduced in
Refs.~\cite{Pagura:2016pwr,GomezDumm:2017iex}, where both zero and finite
temperature cases have been considered. Noticeably, in these articles it is
shown that nlNJL models naturally allow to reproduce the effect of inverse
magnetic catalysis (IMC) observed from LQCD results---that is, the fact that
the chiral restoration critical temperature turns out to be a decreasing
function of $B$. In fact, the observation of IMC in LQCD
calculations~\cite{Bali:2011qj,Bali:2012zg} represents a challenge from the
point of view of theoretical models, since most naive effective approaches
to low energy QCD (NJL model, chiral perturbation theory, MIT bag model,
quark-meson models) predict that the chiral transition temperature should
grow when the magnetic field is
increased~\cite{Andersen:2014xxa,Kharzeev:2012ph,Miransky:2015ava}. As
shown in Refs.~\cite{Ayala:2014iba,Farias:2014eca}, this problem can be
overcome (e.g.~in the case of the local NJL model) by allowing for a $B$
dependence in the coupling constants. In the present paper we show that
nlNJL models not only provide a natural description of the IMC effect but
also lead to a $B$ dependence of the $\pi^0$ mass that is found to be in
good agreement with LQCD results.

This article is organized as follows. In Sec.~II we show how to obtain the
analytical equations required to determine the values of the $\pi^0$ mass
and decay constant in the presence of the magnetic field. Our calculations
are based on the formalism developed in
Refs.~\cite{Pagura:2016pwr,GomezDumm:2017iex}, which makes use of Ritus
eigenfunctions~\cite{Ritus:1978cj}. From this analysis it is also immediate
to obtain an equation for the $\sigma$ scalar meson mass. In the last
subsection of Sec.~II we prove within our model the validity of the
Goldberger-Treiman and Gell-Mann-Oakes-Renner relations in the presence of
the magnetic field. Previous checks of these relations have been carried out
in Refs.~\cite{Agasian:2001ym} and~\cite{Orlovsky:2013wjd} in the framework
of ChPT and PIH, respectively. In Sect.~III we quote and discuss our
numerical results, comparing our findings with those obtained in LQCD. Our
conclusions are presented in Sec.~IV. Finally, in Appendices A and B we
outline the derivation of some expressions quoted in the main text.


\section{Theoretical formalism}

Let us start by stating the Euclidean action for our nonlocal NJL-like
two-flavor quark model,
\begin{equation}
S_E = \int d^4 x \ \left\{ \bar \psi (x) \left(- i \rlap/\partial
+ m_c \right) \psi (x) -
\frac{G}{2} j_a(x) j_a(x) \right\} \ .
\label{action}
\end{equation}
Here $m_c$ is the current quark mass, which is assumed to be equal for $u$
and $d$ quarks. The currents $j_a(x)$ are given by
\begin{eqnarray}
j_a (x) &=& \int d^4 z \  {\cal G}(z) \
\bar \psi(x+\frac{z}{2}) \ \Gamma_a \ \psi(x-\frac{z}{2}) \ ,
\label{cuOGE}
\end{eqnarray}
where $\Gamma_{a}=(\leavevmode\hbox{\small1\kern-3.8pt\normalsize1},i\gamma
_{5}\vec{\tau})$, and the function ${\cal G}(z)$ is a nonlocal form factor
that characterizes the effective interaction. We introduce now in the
effective action Eq.~(\ref{action}) a coupling to an external
electromagnetic gauge field $\mathcal{A}_{\mu}$. 
For a local theory this can be done by performing the replacement
\begin{equation}
\partial_{\mu}\ \rightarrow\ D_\mu\equiv\partial_{\mu}-i\,\hat Q
\mathcal{A}_{\mu}(x)\ ,
\label{covdev}
\end{equation}
where $\hat Q=\mbox{diag}(q_u,q_d)$, with $q_u=2e/3$, $q_d = -e/3$, is the
electromagnetic quark charge operator. In the case of the nonlocal model
under consideration, the inclusion of gauge interactions implies a change
not only in the kinetic terms of the Lagrangian but also in the nonlocal
currents in Eq.~(\ref{cuOGE}). One has
\begin{equation}
\psi(x-z/2) \rightarrow \mathcal{W}\left(  x,x-z/2\right)  \ \psi(x-z/2)\ ,
\label{transport}
\end{equation}
and a related change holds for $\bar
\psi(x+z/2)$~\cite{GomezDumm:2006vz,Noguera:2008,Dumm:2010hh}. Here the
function $\mathcal{W}(s,t)$ is defined by
\begin{equation}
\mathcal{W}(r,s)\ =\ \mathrm{P}\;\exp\left[ -\, i \int_{r}^{s}d\ell_{\mu}\, 
\hat Q\mathcal{A}_{\mu}(\ell)
\right]  \ ,
\label{intpath}%
\end{equation}
where $r$ runs over an arbitrary path connecting $r$ with $s$. As it is
usually done, we take it to be a straight line path.

Since we are interested in studying light meson properties, it is convenient
to bosonize the fermionic theory, introducing scalar and pseudoscalar fields
$\sigma(x)$ and $\vec{\pi}(x)$ and integrating out the fermion fields. The
bosonized action can be written as~\cite{Noguera:2008,Dumm:2010hh}
\begin{equation}
S_{\mathrm{bos}}=-\log\det\mathcal{D}+\frac{1}{2G}
\int d^{4}x
\Big[\sigma(x)\sigma(x)+ \vec{\pi}(x)\cdot\vec{\pi}(x)\Big]\ ,
\label{sbos}
\end{equation}
with
\begin{eqnarray}
\mathcal{D}
\left( x,x' \right)   &  = & \delta^{(4)}(x-x')\,\big(-i\,\rlap/\!D + m_{c} \big)\,
+ \nonumber \\
& & \mathcal{G}(x-x') \, \gamma_{0} \, W(x,\bar x)\,
\gamma_{0} \big[\sigma(\bar x) + i\,\gamma_5\,\vec{\tau}\cdot\vec{\pi}(\bar x) \big]
\, W(\bar x,x') \ ,
\label{dxx}%
\end{eqnarray}
where we have defined $\bar x = (x+x')/2$. We will consider the particular
case of a constant and homogenous magnetic field orientated along the
positive direction of the 3 axis. Then, in the Landau gauge, one has
$\mathcal{A}_\mu = B\, x_1\, \delta_{\mu 2}$.

\subsection{Mean field fermion propagator}

We proceed by expanding the operator in Eq.~(\ref{dxx}) in powers of the
fluctuations $\delta\pi_i$ and $\delta\sigma$ around the corresponding mean
field values. We assume that the field $\sigma(x)$ has a nontrivial
translational invariant mean field value $\bar{\sigma}$, while the vacuum
expectation values of pseudoscalar fields are zero. Thus we write
\begin{equation}
\mathcal{D}(x,x') = \mathcal{D}^{\mbox{\tiny MFA}}(x,x') +
\delta\mathcal{D}(x,x')\ .
\end{equation}
It is easy to see that the mean field piece is flavor diagonal. One has
\begin{equation}
\mathcal{D}^\mfa(x,x') \ = \ {\rm diag}\big(\mathcal{D}_u^\mfa(x,x')\, ,\,
\mathcal{D}_d^\mfa(x,x')\big)\ ,
\end{equation}
where
\begin{equation}
\mathcal{D}_f^\mfa(x,x') \ = \ \delta^{(4)}(x-x') \left( - i \rlap/\partial
- q_f \, B \, x_1 \, \gamma_2 + m_c \right)  + \, \bar\sigma \,
\mathcal{G}(x-x') \, \exp\left[i\Phi_f(x,x')\right]\ .
\end{equation}
Here a direct product to an identity matrix in color space is understood. It
is seen that the operator $\mathcal{D}_f^\mfa(x,x')$ includes a
translational invariant piece, plus a term carrying the nonlocal form factor
and the so-called Schwinger phase $\Phi_f(x,x')= q_f B \, (x_2 -
x_2')\, (x_1 +x_1')/2\,$. The mean field quark propagators $S_f^\mfa(x,x')$ are
defined now as
\begin{equation}
S_f^\mfa(x,x') \ = \ \big[\mathcal{D}_f^\mfa(x,x')\big]^{-1}\ .
\end{equation}
Their explicit form can be obtained by following the Ritus eigenfunction
method~\cite{Ritus:1978cj}. As shown in Ref.~\cite{GomezDumm:2017iex} (see
also the analysis carried out within the Schwinger-Dyson formalism in
Refs.~\cite{Watson:2013ghq,Mueller:2014tea}), the propagators can be written
in terms of the Schwinger phase $\Phi_f(x,x')$ and a translational invariant
function, namely
\begin{equation}
S_f^\mfa(x,x') \ = \ \exp\!\big[i\Phi_f(x,x')\big]\,\int \frac{d^4p}{(2\pi)^4}\
e^{i\, p\cdot (x-x')}\, \tilde S_f(p_\perp,p_\parallel)\ ,
\end{equation}
where $p_\perp = (p_1,p_2)$ and $p_\parallel = (p_3,p_4)$. The
expression of $\tilde S_f(p_\perp,p_\parallel)$ in the nlNJL model is found
to be~\cite{GomezDumm:2017iex}
\begin{eqnarray}
\tilde S_f(p_\perp,p_\parallel) & = & 2\, \exp(-p_\perp^2/|q_f B|)
\sum_{k=0}^\infty \sum_{\lambda=\pm} \Big[(-1)^{k_\lambda}
\big(\hat A^{\lambda,f}_{k,p_\parallel} - \hat B^{\lambda,f}_{k,p_\parallel}
\, p_\parallel\cdot\gamma_\parallel\big) L_{k_\lambda}(2p_\perp^2/|q_f B|)
+ \nonumber\\
& & 2\, (-1)^k \big(\hat C^{\lambda,f}_{k,p_\parallel}
- \hat D^{\lambda,f}_{k,p_\parallel}\, p_\parallel\cdot\gamma_\parallel\big)
\, p_\perp\cdot\gamma_\perp
\, L^1_{k-1}(2p_\perp^2/|q_f B|)\Big]\,\Delta^\lambda\ ,
\label{sfp}
\end{eqnarray}
where the following definitions have been used. The perpendicular and
parallel gamma matrices are collected in vectors $\gamma_\perp =
(\gamma_1,\gamma_2)$ and $\gamma_\parallel = (\gamma_3,\gamma_4)$, while the
matrices $\Delta^\lambda$ are defined as $\Delta^+ = {\rm diag}(1,0,1,0)$
and $\Delta^- = {\rm diag}(0,1,0,1)$. The integers $k_\lambda$ are given by
$k_\pm = k - 1/2 \pm s_f/2$, where $s_f = {\rm sign} (q_f B)$. The functions
$\hat X^{\pm,f}_{k,p_\parallel}$, with $X = A, B, C, D$, are defined as
\begin{eqnarray}
\hat A^{\pm,f}_{k,p_\parallel} & = &
M^{\mp,f}_{k,p_\parallel}\, \hat C^{\pm,f}_{k,p_\parallel} + p_\parallel^2
\, \hat D^{\pm,f}_{k,p_\parallel}\ ,
\label{aa} \\
\hat B^{\pm,f}_{k,p_\parallel} & = & \hat C^{\pm,f}_{k,p_\parallel}
- M^{\mp,f}_{k,p_\parallel}\, \hat D^{\pm,f}_{k,p_\parallel}\ \ ,
\label{bb} \\
\hat C^{\pm,f}_{k,p_\parallel} & = & \frac{2 k |q_f B| + p_\parallel^2 +
M^{-,f}_{k,p_\parallel} M^{+,f}_{k,p_\parallel}}{\Delta^f_{k,p_\parallel}}
\ \ ,
\label{cc} \\
\hat D^{\pm,f}_{k,p_\parallel} & = & \frac{M^{\pm,f}_{k,p_\parallel}
- M^{\mp,f}_{k,p_\parallel}}{\Delta^f_{k,p_\parallel}} \ \ ,
\label{dd}
\end{eqnarray}
where
\begin{equation}
\Delta^f_{k,p_\parallel} = \left( 2 k |q_f B| + p_\parallel^2 +
M^{+,f}_{k,p_\parallel}\, M^{-,f}_{k,p_\parallel} \right)^2\! +\, p_\parallel^2
\left( M^{+,f}_{k,p_\parallel} - M^{-,f}_{k,p_\parallel} \right)^2\ ,
\end{equation}
whereas the functions $M^{\lambda,f}_{k,p_\parallel}$ play the role of
effective (momentum-dependent) dynamical quark masses in presence of the
magnetic field. They are given by
\begin{equation}
M^{\lambda,f}_{k,p_\parallel} \ = \
\frac{4\pi}{|q_fB|}\,(-1)^{k_\lambda}
\int \frac{d^2p_\perp}{(2\pi)^2}\
M(p_\perp^2 + p_\parallel^2) \,\exp(-p_\perp^2/|q_fB|) \,
L_{k_\lambda}(2p_\perp^2/|q_fB|)\ ,
\label{mpk}
\end{equation}
where
\begin{equation}
M(p^2) \ = \ m_c + \bar\sigma\,g(p^2)\ ,
\end{equation}
$g(p^2)$ being the Fourier transform of the nonlocal form factor ${\cal
G}(x)$. In Eqs.~(\ref{sfp}) and (\ref{mpk}), $L_k(x)$, $L_k^1(x)$ stand for
generalized Laguerre polynomials, with the convention $L_{-1}(x) =
L_{-1}^1(x) = 0$. The relation in Eq.~(\ref{mpk}) can be understood as a
Laguerre-Fourier transform of the function $M(p^2)$. It is also convenient
to introduce Laguerre-Fourier transforms of the form factor $g(p^2)$,
\begin{equation}
g^{\lambda,f}_{k,p_\parallel} \ = \
\frac{4\pi}{|q_fB|}\,(-1)^{k_\lambda}
\int \frac{d^2p_\perp}{(2\pi)^2}\
g(p_\perp^2 + p_\parallel^2) \,\exp(-p_\perp^2/|q_fB|) \,
L_{k_\lambda}(2p_\perp^2/|q_fB|)\ ,
\label{gpk}
\end{equation}
thus one has
\begin{equation}
M^{\lambda,f}_{k,p_\parallel} \ = \ \big[1-\delta_{(k_\lambda+1)\,0}\big]
m_c\, + \,\bar \sigma \, g^{\lambda,f}_{k,p_\parallel}\ .
\label{mmain}
\end{equation}

Let us also quote the expressions for the quark condensates, $\langle \bar
uu\rangle$ and $\langle \bar dd\rangle$, which can be obtained from
\begin{equation}
\langle \bar ff\rangle \ = \ -\, \frac{1}{V^{(4)}}\, {\rm Tr}\, S_f^\mfa \ = \
-N_C\int \frac{d^4p}{(2\pi)^4}\
{\rm tr}_D\,\tilde S_f(p_\perp,p_\parallel) \ .
\end{equation}
Given the result for the propagators in Eq.~(\ref{sfp}) one gets
\begin{eqnarray}
\langle \bar ff\rangle & = & -\,4
\sum_{k=0}^\infty \int_{q_\parallel}\sum_{\lambda=\pm} (-1)^{k_\lambda}
\hat A^{\lambda,f}_{k,p_\parallel} \int \frac{d^2p_\perp}{(2\pi)^2}
\, \exp(-p_\perp^2/|q_f B|)\, L_{k_\lambda}(2p_\perp^2/|q_f B|) \nonumber \\
& = & -\,\frac{N_C\, |q_f B|}{\pi}
\sum_{k=0}^\infty\; \int_{q_\parallel}
\sum_{\lambda = \pm}\, \hat A_{k,q_\parallel}^{\lambda,f}\ .
\label{conden}
\end{eqnarray}
As usual in this type of models, it is seen that the chiral condensates
turn out to be divergent away from the chiral limit, thus they have to be
regularized. We follow here a prescription similar as that considered e.g.\
in Ref.~\cite{GomezDumm:2004sr}, in which we subtract the corresponding free
quark contribution and then we add it in a regularized form. Thus we have
\begin{equation}
\langle \bar ff\rangle_{\rm reg} \ = \ \langle \bar ff\rangle
\, - \langle \bar ff\rangle_{\rm free} + \langle \bar ff\rangle_{\rm
free,reg}\ ,
\end{equation}
where the free, regularized contribution (notice that by ``free'' we mean in
absence of the four fermion effective coupling, but keeping the interaction
with the magnetic field) is given
by~\cite{Ebert:1999ht,Menezes:2008qt,GomezDumm:2017iex}
\begin{equation}
\langle \bar ff\rangle_{\rm free,reg}(B)\ = \
 \frac{N_C\, m_c^3}{4 \pi^2} \left[ \frac{\ln \Gamma(x_f)}{x_f} - \frac{\ln 2
\pi}{2 x_f} + 1 - \left(1-\frac{1}{2x_f}\right) \ln x_f \right]\ ,
\end{equation}
with $x_f = m_c^2/(2 |q_f B|)$.

\subsection{$\pi^0$ and $\sigma$ meson masses}

The expression of the quark propagator in Eq.~(\ref{sfp}) can be used to
obtain the theoretical expressions for the $\pi^0$ and $\sigma$ meson masses
within the nlNJL model. Let us first concentrate on the $\pi^0$ mass, which
follows from the terms in the expansion of the bosonized action $S_{\rm
bos}$ that are quadratic in $\delta\pi_3$. Expanding the first term in
Eq.~(\ref{sbos}) around the mean field values of the meson fields one has
\begin{eqnarray}
-\log\det\mathcal{D} & = & - \tra\log\mathcal{D}_0 \, - \, \tra\log(1 +
\mathcal{D}_0^{-1} \,\delta\mathcal{D}) \nonumber \\
& = & - \tra\log\mathcal{D}_0 \, - \,
\tra(\mathcal{D}_0^{-1}\,\delta\mathcal{D}) \, + \,
\frac{1}{2}\,\tra(\mathcal{D}_0^{-1}\,\delta\mathcal{D})^2 \, + \, \dots
\label{expansion}
\end{eqnarray}
{}From Eq.~(\ref{dxx}), it is seen that the quadratic piece is given by
\begin{eqnarray}
\frac{1}{2}\,\tra(\mathcal{D}_0^{-1}\,\delta\mathcal{D})^2\bigg|_{(\delta\pi_3)^2}
& = & - \frac{1}{2} \int \mathcal{G}(x'-x'')\,\mathcal{G}(x'''-x) \;
\trmin_{cfD} \bigg[ \mathcal{D}_0^{-1}(x,x') \, \gamma_5
\, \exp[\Phi(x',x'')] \nonumber \\
& & \hspace{-1.5cm}\times\, \mathcal{D}_0^{-1}(x'',x''')\, \gamma_5
\, \exp[\Phi(x''',x)]\bigg]\,
\delta\pi_3\Big(\frac{x'+x''}{2}\Big)\;
\delta\pi_3\Big(\frac{x'''+x}{2}\Big)\ ,
\label{tracepipi}
\end{eqnarray}
where the integral extends over coordinate spaces $x$, $x'$, $x''$ and
$x'''$, and the trace acts on color, flavor and Dirac spaces.

To determine the $\pi^0$ mass it is convenient to write the trace in
Eq.~(\ref{tracepipi}) in momentum space. In this way the $(\delta\pi_3)^2$
piece of the bosonized action in Eq.~(\ref{sbos}) can be written as
\begin{eqnarray}
S_{\rm bos}\big|_{(\delta\pi_3)^2} & = &
\frac{1}{2}\,\tra(\mathcal{D}_0^{-1}\,\delta\mathcal{D})^2\Big|_{(\delta\pi_3)^2}
+ \frac{1}{2G}\int \frac{d^4t}{(2\pi)^4}\ \delta\pi_3(t)\,\delta\pi_3(-t)
\nonumber \\
& = & \frac{1}{2}\int \frac{d^4t}{(2\pi)^4} \left[F(t_\perp^2,t_\parallel^2) + \frac{1}{G}\right]
\delta\pi_3(t)\,\delta\pi_3(-t)\ ,
\end{eqnarray}
and, choosing the frame in which the $\pi^0$ meson is at rest, its mass can be
obtained as the solution of the equation
\begin{equation}
F(0,-m_{\pi^0}^2) +\frac{1}{G}\ = \ 0\ .
\label{pimass}
\end{equation}
Thus, our task is to obtain within our model the function
$F(t_\perp^2,t_\parallel^2)$ in the limit $t_\perp = 0$. After some
straightforward calculation, from Eq.~(\ref{tracepipi}) one gets
\begin{eqnarray}
F(0,t_\parallel^2) & = & 16\,\pi^2\,N_C\, \sum_{f=u,d}\,\frac{1}{(q_f B)^2}
\int_{q_\perp\, p_\perp\, p'_\perp\, q_\parallel} g(q_\perp^2 +
q_\parallel^2)\, g[(p'_\perp + p_\perp - q_\perp)^2\! + q_\parallel^2]\,\times
\nonumber \\
& & \exp[i2\phi(q_\perp,p_\perp,p'_\perp)/(q_f B)] \ \trmin_D \Big[
\tilde S_f(p_\perp,q_\parallel^+)\,i\gamma_5\,
\tilde S_f(p'_\perp,q_\parallel^-)\,i\gamma_5\Big]\ ,
\label{f0k}
\end{eqnarray}
where we have defined $q_\parallel^\pm = q_\parallel \pm t_\parallel/2$, and the
function $\phi$ in the exponential is given by
\begin{equation}
\phi(q_\perp,p_\perp,p'_\perp) \ = \ p_2\,p'_1 + q_1\,(p'_2 - p_2)
- p_1\,p'_2 - q_2\,(p'_1 - p_1) \ .
\label{phi}
\end{equation}
For the integrals over two-dimensional momentum vectors we have used the notation
\begin{equation}
\int_{p\, q\, \dots} \ \equiv \ \int \frac{d^2p}{(2\pi)^2}\; \frac{d^2q}{(2\pi)^2}\ \dots
\end{equation}
The evaluation of the trace in Eq.~(\ref{f0k}) leads to
\begin{eqnarray}
\trmin_D \Big[
\tilde S_f(p_\perp,q_\parallel^+)\,i\gamma_5\,
\tilde S_f(p'_\perp,q_\parallel^-)\,i\gamma_5\Big] & = &
-\, 8\,e^{-(p_\perp^2+{p'_\perp}^{\!\! 2})/B_f}\sum_{k,k'=0}^\infty
(-1)^{k+k'}\,\times
\nonumber \\
& & \hspace{-5.4cm} \bigg[\sum_{\lambda = \pm}
F_{kk',q_\parallel^+q_\parallel^-}^{\lambda,f\,(AB)}\,
L_{k_\lambda}(2p_\perp^2/B_f)\, L_{k_\lambda'}(2{p'_\perp}^{\!\!2}/B_f) \ + \nonumber \\
& & \hspace{-5.1cm}
8\, F_{kk',q_\parallel^+q_\parallel^-}^{+,f\,(CD)}\,(p\cdot p')\,
L_{k-1}^1(2p_\perp^2/B_f)\, L_{k'-1}^1(2{p'_\perp}^{\!\!2}/B_f) \bigg]\ ,
\label{tracess}
\end{eqnarray}
where
\begin{equation}
F_{kk',q_\parallel^+q_\parallel^-}^{\lambda,f\,(XY)} \ = \ \hat X^{\lambda,f}_{k,q_\parallel^+}
\,\hat X^{\lambda,f}_{k',q_\parallel^-} + (q_\parallel^+ \cdot
q_\parallel^-)\, \hat Y^{\lambda,f}_{k,q_\parallel^+}\,\hat Y^{\lambda,f}_{k',q_\parallel^-}\ .
\label{fxy}
\end{equation}
For simplicity we have introduced here the notation $B_f = |q_f B|$.

To work out the integrals over $p_\perp$, $p_\perp'$ and $q_\perp$, which
involve the Laguerre polynomials, it is convenient to introduce the
Laguerre-Fourier transforms of the nonlocal form factors. It is seen that
Eq.~(\ref{gpk}) can be inverted to get
\begin{equation}
g(p_\perp^2 + p_\parallel^2) \ = \ 2\, e^{-p_\perp^2/B_f}\;
\sum_{k=0}^\infty\; (-1)^{k_\lambda}\,
g^{\lambda,f}_{k,p_\parallel}\,L_{k_\lambda}(2p_\perp^2/B_f)\ ,
\label{inversa}
\end{equation}
for either $\lambda = +$ or $\lambda = -$. Using this relation to transform
the functions $g(q_\perp^2 + q_\parallel^2)$ and $g[(p'_\perp + p_\perp -
q_\perp)^2+q_\parallel^2]$ in Eq.~(\ref{f0k}), it can be shown that the
integrals over perpendicular momenta can be performed analytically. The
corresponding calculation, sketched in Appendix A, leads to a relatively
brief expression for $F(0,t_\parallel^2)$, namely
\begin{equation}
F(0,t_\parallel^2) = -\frac{N_C}{\pi} \sum_{f=u,d} B_f\sum_{k=0}^\infty\int
\frac{d^2q_\parallel}{(2\pi)^2} \bigg[\sum_{\lambda = \pm}
{g^{\lambda,f}_{k,q_\parallel}}^2 \,
F_{kk,q_\parallel^+q_\parallel^-}^{\lambda,f\,(AB)} +
4k\,B_f\,g^{+,f}_{k,q_\parallel}\,g^{-,f}_{k,q_\parallel}\,
F_{kk,q_\parallel^+q_\parallel^-}^{+,f\,(CD)}
\bigg] \, ,
\label{f0kfinal}
\end{equation}
which is one of the main analytical results of this article. In the limit
$B\to 0$, it can be shown that Eq.~(\ref{f0kfinal}) reduces, as it should,
to the expression quoted e.g.\ in Ref.~\cite{GomezDumm:2006vz},
\begin{equation}
F(t^2)\bigg|_{B=0} \ = \ -\,8N_C\,\int\ \frac{d^4q}{(2\pi)^4}\ g(q^2)^2\,\frac{(q^+\cdot
q^-)\, +\, M({q^+}^2)\,M({q^-}^2)}{\left[{q^+}^2 + M({q^+}^2)\right]
\left[{q^-}^2 + M({q^-}^2)\right]}\,\ .
\label{fb0}
\end{equation}

In the case of the $\sigma$ meson, the mass can be determined from a
relation similar to Eq.~(\ref{pimass}). The corresponding function
$G(0,t_\parallel^2)$ is obtained by following basically the same steps as
for the $\pi^0$ case. The essential difference is that one has to remove the
factors $i\gamma_5$ in the trace in Eq.~(\ref{f0k}). When calculating this
trace one arrives at a result analogous to that in Eq.~(\ref{tracess}),
where the new functions
$G_{kk',q_\parallel^+q_\parallel^-}^{\lambda,f\,(XY)}$ are given by
\begin{eqnarray}
G_{kk',q_\parallel^+q_\parallel^-}^{\lambda,f\,(AB)} & = & - \hat A^{\lambda,f}_{k,q_\parallel^+}
\,\hat A^{\lambda,f}_{k',q_\parallel^-} + (q_\parallel^+ \cdot
q_\parallel^-)\, \hat B^{\lambda,f}_{k,q_\parallel^+}\,\hat
B^{\lambda,f}_{k',q_\parallel^-}\nonumber \\
G_{kk',q_\parallel^+q_\parallel^-}^{\lambda,f\,(CD)} & = & \hat C^{\lambda,f}_{k,q_\parallel^+}
\,\hat C^{\lambda,f}_{k',q_\parallel^-} - (q_\parallel^+ \cdot
q_\parallel^-)\, \hat D^{\lambda,f}_{k,q_\parallel^+}\,\hat
D^{\lambda,f}_{k',q_\parallel^-}\ .
\label{gxy}
\end{eqnarray}
The final expression for $G(0,t_\parallel^2)$ has then the same form as the
lhs of Eq.~(\ref{f0kfinal}), just replacing
$F_{kk',q_\parallel^+q_\parallel^-}^{\pm,f\,(XY)} \to
G_{kk',q_\parallel^+q_\parallel^-}^{\pm,f\,(XY)}$.

\subsection{$\pi^0$ decay constant}

The $\pi^0$ decay constant is defined through the matrix element of the
axial current ${\cal J}_{A3}^\mu$ between the vacuum and the physical pion
state, taken at the pion pole. One has
\begin{equation}
\langle 0|\,{\cal J}^\mu_{A3}(x)\,|\tilde \pi_3(t)\rangle \ = \
i\,e^{-i(t\cdot x)}\,f(t^2)\, t^\mu\ ,
\label{fpidef}
\end{equation}
where $\tilde \pi_3(t) = Z_{\pi^0}^{-1/2}\pi_3(t)$ is the renormalized field
associated with the $\pi^0$ meson state, with $t^2 = -m_{\pi^0}^2$. In the
presence of the external magnetic field, the wave function renormalization
factor $Z_{\pi^0}^{1/2}$ is given by the residue of the pion propagator at
$t^2 = -m_{\pi^0}^2$, i.e.
\begin{equation}
Z_{\pi^0}^{-1} \ = \
\frac{dF(0,t_\parallel^2)}{dt_\parallel^2}\bigg|_{t_\parallel^2 =
-m_{\pi^0}^2}\ ,
\label{zpi}
\end{equation}
where $F(0,t_\parallel^2)$ is the function in Eq.~(\ref{f0kfinal}). The
matrix element in Eq.~(\ref{fpidef}) can be obtained by introducing a
coupling between the current ${\cal J}^\mu_{A3}$ and an auxiliary axial
gauge field $W_3^\mu$, and taking the corresponding functional derivative of
the effective action. In the same way as discussed at the beginning of this
section, gauge invariance requires the couplings to this auxiliary gauge
field to be introduced through the covariant derivative and the parallel
transport of the fermion fields, see Eqs.~(\ref{covdev}) and
(\ref{transport}). In the presence of the external magnetic field one has
\begin{eqnarray}
D_\mu & = & \partial_\mu - i\,\hat Q
\mathcal{A}_{\mu}(x) -\,\frac{i}{2}\,\gamma_5\,\tau_3 W_{3\mu}(x)\ , \\
\mathcal{W}(r,s) & = & \mathrm{P}\;\exp\bigg\{ -\, i \int_{r}^{s}d\ell_{\mu}\,
\left[ \hat Q\mathcal{A}_{\mu}(\ell)
 + \frac{1}{2}\,\gamma_5\,\tau_3 W_{3\mu}(\ell)\right]\bigg\} \ .
\end{eqnarray}
Assuming that the mean field value of the $\pi_3$ field vanishes, the pion
decay constant can be obtained by expanding the bosonized action up to first
order in $W_{3\mu}$ and $\delta\pi_3$. Writing
\begin{equation}
S_{\rm bos}\big|_{W_3\,\delta\pi_3} \ = \ \int \frac{d^4t}{(2\pi)^4}
\, F_\mu(t)\; W_{3\mu}(t)\;\delta\pi_3(-t)\ ,
\end{equation}
one finds
\begin{equation}
f_{\pi^0} \ = \ f(-m_{\pi^0}^2) \ = \ i\,Z_{\pi^0}^{1/2}\,\frac{t_\mu
F_\mu(t)}{t^2}\bigg|_{t_\perp^2=0,t_\parallel^2 = -m_{\pi^0}^2}\ .
\label{fpi}
\end{equation}

To find the function $F_\mu(t)$ we consider once again the expansion in
Eq.~(\ref{expansion}). In addition, we expand $\delta{\cal D}$ in powers of
$\delta \pi_3$ and $W_3$,
\begin{equation}
\delta\mathcal{D} \ = \ \delta\mathcal{D}_{W} + \delta\mathcal{D}_\pi +
\delta\mathcal{D}_{W\pi} +\ \dots\ ,
\label{deltadexp}
\end{equation}
which leads to
\begin{equation}
S_{\rm bos}\big|_{W_3\,\delta\pi_3} \ = \
-\,\tra(\mathcal{D}_0^{-1}\,\delta\mathcal{D}_{W\pi})\, +
\,\tra(\mathcal{D}_0^{-1}\,\delta\mathcal{D}_{W}\mathcal{D}_0^{-1}\,
\delta\mathcal{D}_\pi)\ .
\label{sbosfpi}
\end{equation}
The operators in the rhs of Eq.~(\ref{deltadexp}) explicitly read
\begin{eqnarray}
\delta\mathcal{D}_\pi(x,x') & = & i\,\gamma_5\,\tau_3\,\exp[\Phi(x,x')]\,
\mathcal{G}(x-x')\,\delta\pi_3(\bar x)\ ,\\
\delta\mathcal{D}_{W}(x,x') & = & \delta^{(4)}(x-x')\,
\frac{\tau_3}{2}\,\gamma_5\,\gamma_\mu\,W_{3\mu}(\bar x)\, + \nonumber \\
& & i\,\bar\sigma\,\gamma_5\,\frac{\tau_3}{2}\,\exp[\Phi(x,x')]\,
\mathcal{G}(x-x')\,\big[a_3(x,\bar x)-a_3(\bar x,x)\big]\ , \
\label{dw} \\
\delta\mathcal{D}_{W\pi}(x,x') & = &
-\,\frac{1}{2}\,\exp[\Phi(x,x')]\,\mathcal{G}(x-x')
\,\big[a_3(x,\bar x)-a_3(\bar x,x)\big]\,\delta\pi_3(\bar x)\ ,
\end{eqnarray}
where we have introduced the definitions $\bar x = (x+x')/2$ and
\begin{equation}
a_3(x,y) \ = \ \int_x^y\ d\ell_\mu\, W_{3\mu}(\ell)\ .
\end{equation}
A direct product to an identity matrix in color space is understood.

The first and second terms in the rhs of Eq.~(\ref{sbosfpi}) can be
diagrammatically represented as a tadpole and a two-propagator contribution,
respectively. Let us start by discussing the tadpole piece. After some
straightforward calculation we get
\begin{equation}
-\,\tra(\mathcal{D}_0^{-1}\,\delta\mathcal{D}_{W\pi}) \ = \ \int
\frac{d^4t}{(2\pi)^4}\; F^{({\rm I})}_\mu(t)\; W_{3\mu}(t)\;\delta\pi_3(-t)\ ,
\end{equation}
where
\begin{eqnarray}
F^{({\rm I})}_\mu(t) & = & i\,\frac{N_C}{2}\sum_{f=u,d}\int\,
\frac{d^4p}{(2\pi)^4}\,\frac{d^4q}{(2\pi)^4}\; \Big\{
g\big[(p-q/2)^2\big] - g\big[(p-q/2+t/2)^2\big]\Big\}\,\times
\nonumber \\
& & \trmin_D \Big[\tilde S_f(p_\perp,p_\parallel)\Big]\, h_\mu(q,t-q)\ ,
\end{eqnarray}
with
\begin{equation}
h_\mu(q,q') \ = \ -\,i\int\, d^4z\ e^{i q'\cdot z}\int_0^z d\ell_\mu\;
e^{i(q+q')\cdot \ell}\ .
\label{hmu}
\end{equation}
Since we are interested in the scalar product $t\cdot F^{(I)}(t)$, we can
use the relation
\begin{equation}
t_\mu\,h_\mu(q,t-q) \ = \
(2\pi)^4\big[\,\delta^{(4)}(t-q)-\delta^{(4)}(q)\,\big]\ ,
\label{relation}
\end{equation}
which holds independently of the integration path chosen in Eq.~(\ref{hmu}).
Taking into account the expression for $\tilde S_f(p_\perp,p_\parallel)$ in
Eq.~(\ref{sfp}) we obtain
\begin{eqnarray}
k_\mu\,F^{({\rm I})}_\mu(t)\Big|_{t_\perp=0} & = & i\, 2\, N_C
\sum_{f=u,d}\;\sum_{k=0}^\infty\; \int_{p_\perp\,p_\parallel}\,
\big[g({p^+}^2) + g({p^-}^2) - 2\,g(p^2)\big]\,\times
 \nonumber \\
& &
\exp(-p_\perp^2/B_f) \sum_{\lambda = \pm} (-1)^{k_\lambda}
\hat A_{k,p_\parallel}^{\lambda,f}\,
L_{k_\lambda}(2p_\perp^2/B_f)\ ,
\end{eqnarray}
where ${p^\pm}^2 = p_\perp^2 + (p_\parallel \pm t_\parallel/2)^2$. Now, as
in the case of the meson masses, we can perform the integral over $p_\perp$
after taking the Laguerre-Fourier transform of the nonlocal form factors,
Eq.~(\ref{inversa}). We have
\begin{eqnarray}
t_\mu\,F^{({\rm I})}_\mu(t)\Big|_{t_\perp=0} & = & i\,4N_C
\sum_{f=u,d}\,\sum_{k,k'=0}^\infty(-1)^{k+k'}\int_{p_\perp p_\parallel}
\exp(-2p_\perp^2/B_f)\,\times
\nonumber \\
& & \sum_{\lambda = \pm}
\,\Big(g^{\lambda,f}_{k',p_\parallel^+} + g^{\lambda,f}_{k',p_\parallel^-}
-2\,g^{\lambda,f}_{k',p_\parallel}\Big)\,\hat A_{k,p_\parallel}^{\lambda,f}\,
L_{k'_\lambda}(2p_\perp^2/B_f)\,
L_{k_\lambda}(2p_\perp^2/B_f)
\nonumber \\
& = & \,i\,\frac{N_C}{2\pi}
\sum_{f=u,d} B_f \sum_{k=0}^\infty\,\int_{p_\parallel}
\sum_{\lambda = \pm} \,
\Big(g^{\lambda,f}_{k,p_\parallel^+} + g^{\lambda,f}_{k,p_\parallel^-}
-2\,g^{\lambda,f}_{k,p_\parallel}\Big)\,\hat A_{k,p_\parallel}^{\lambda,f}
\ ,
\label{f1}
\end{eqnarray}
where we have made use of the orthogonality property of Laguerre
polynomials.

To analyze the two-propagator piece we write
\begin{equation}
\tra(\mathcal{D}_0^{-1}\,\delta\mathcal{D}_{W}\mathcal{D}_0^{-1}\,
\delta\mathcal{D}_\pi) \ = \ \int\
\frac{d^4t}{(2\pi)^4} \Big[F^{({\rm II})}_\mu(t) + F^{({\rm III})}_\mu(t)\Big]
\; W_{3\mu}(t)\;\delta\pi_3(-t)\ ,
\label{f2masf3}
\end{equation}
where $F^{({\rm II})}_\mu(t)$ and $F^{({\rm III})}_\mu(t)$ correspond to the
contributions arising from the first and second terms of
$\delta\mathcal{D}_W$ in Eq.~(\ref{dw}), respectively. For the first term we
obtain
\begin{eqnarray}
F^{({\rm II})}_\mu(t) & = &
i\,8\pi^2N_C\sum_{f=u,d}\frac{1}{B_f^2}\int_{q_\parallel\,q_\perp\,p_\perp\,p'_\perp}
g(q^2) \,\exp\big[i2\varphi(q_\perp,p_\perp,p'_\perp,t_\perp)/(q_f B)\big]\,\times
\nonumber \\
& &
\trmin_D \Big[\tilde S_f(p_\perp,q_\parallel^+)\,\gamma_5\,\gamma_\mu\,
\tilde S_f(p'_\perp,q_\parallel^-)\,\gamma_5\Big]\ ,
\label{f2ini}
\end{eqnarray}
where $q_\parallel^\pm = q_\parallel\pm t_\parallel/2$, and the function
$\varphi$ in the exponential is given by
\begin{equation}
\varphi(q_\perp,p_\perp,p'_\perp,t_\perp) \ = \ p_2\,(q_1-t_1/2) -
p_2'\,(q_1+t_1/2)-q_1\,t_2 - p_2\,p'_1 - (1\leftrightarrow 2)\ .
\label{varphi}
\end{equation}
Since we are interested in the product $t_\mu\, F^{({\rm II})}_\mu(t)$ for
$t_\perp = 0$, we calculate the trace
\begin{eqnarray}
\trmin_D \Big[
\tilde S_f(p_\perp,q_\parallel^+)\,\gamma_5\,(t_\parallel\cdot
\gamma_\parallel)\,\tilde S_f(p'_\perp,q_\parallel^-)\,\gamma_5\Big] & = &
8\,\exp\big[-\!(p_\perp^2+{p'_\perp}^{\!\! 2})/B_f\big]\sum_{k,k'=0}^\infty
(-1)^{k+k'}\,\times
\nonumber \\
& & \!\!\!\hspace{-5.9cm} \bigg\{\sum_{\lambda = \pm}\Big[
(t_\parallel \cdot q_\parallel^-)\,\hat A_{k,q_\parallel^+}^{\lambda,f}
\,\hat B_{k',q_\parallel^-}^{\lambda,f} -
(t_\parallel \cdot q_\parallel^+)\,\hat A_{k',q_\parallel^-}^{\lambda,f}
\,\hat B_{k,q_\parallel^+}^{\lambda,f}\Big]
L_{k_\lambda}(2p_\perp^2/B_f)\, L_{k'_\lambda}(2{p'_\perp}^{\!\!2}/B_f) \ + \nonumber \\
& & \!\!\!\hspace{-5.5cm} 8\,i\,(p_1p'_2-p_2p'_1)\,\Big[ (t_\parallel \cdot q_\parallel^-)
\,\hat C_{k,q_\parallel^+}^{+,f} \,\hat D_{k',q_\parallel^-}^{+,f} -
(t_\parallel \cdot q_\parallel^+)\,\hat C_{k',q_\parallel^-}^{+,f}
\,\hat D_{k,q_\parallel^+}^{+,f}\Big] \times \nonumber \\
& & \!\!\!\hspace{-5.5cm} L^1_{k-1}(2p_\perp^2/B_f)\,
L^1_{k-1}(2{p'_\perp}^{\!\!2}/B_f)\bigg\} \ .
\end{eqnarray}
One can now introduce the transformation in Eq.~(\ref{inversa}) for $g(q^2)$
in order to integrate over transverse momenta and express the result in
terms of Laguerre-Fourier transforms of the form factors. This calculation,
outlined in Appendix B, leads to
\begin{eqnarray}
t_\mu\,F^{({\rm II})}_\mu(t)\Big|_{t_\perp=0} & = & -\,i\,\frac{N_C}{\pi}
\sum_{f=u,d}B_f\,\sum_{k=0}^\infty\; \int_{q_\parallel}
\; (t_\parallel\cdot q_\parallel^+)\,\times \nonumber \\
& & \bigg[\, \sum_{\lambda = \pm}\,
g_{k,q_\parallel}^{\lambda,f}\,\hat A_{k,q_\parallel^-}^{\lambda,f}
\,\hat B_{k,q_\parallel^+}^{\lambda,f} + 2\,k\,B_f\,\big(
g_{k,q_\parallel}^{+,f} - g_{k,q_\parallel}^{-,f}\big)\,
\hat C_{k,q_\parallel^-}^{+,f}\,\hat D_{k,q_\parallel^+}^{+,f}\, \bigg]
\ .
\label{f2}
\end{eqnarray}
Finally, for the second term in Eq.~(\ref{f2masf3}) we find
\begin{eqnarray}
F^{({\rm III})}_\mu(t) & = &
i\,8\pi^2\,N_C\;\bar\sigma\, \sum_{f=u,d}\frac{1}{B_f^2}\int \frac{d^4
r}{(2\pi)^4}\;
h_\mu(r,t-r)\,\int_{q_\parallel\,q_\perp\,p_\perp\,p'_\perp}
g(q^2)\,\times\nonumber \\
& & \Big\{ g\big[(p_\perp-r_\perp/2-t_\perp/2)^2 +
(p_\parallel + p'_\parallel - q_\parallel - r_\parallel/2)^2 \big] -
\nonumber \\
& & g\big[(p_\perp-r_\perp/2)^2 +
(p_\parallel + p'_\parallel - q_\parallel - r_\parallel/2 + t_\parallel/2)^2
\big] \Big\}\,\times
\nonumber \\
& & \exp\big[i2\varphi(q_\perp,p_\perp,p'_\perp,k_\perp)/(q_fB)\big]\;
\trmin_D \Big[\tilde S_f(p_\perp,q_\parallel^+)\,i\gamma_5\,
\tilde S_f(p'_\perp,q_\parallel^-)\,i\gamma_5\Big]\ ,
\end{eqnarray}
where the function $\varphi(q_\perp,p_\perp,p'_\perp,k_\perp)$ is that given
in Eq.~(\ref{varphi}). Using the relation in Eq.~(\ref{relation}) we obtain
\begin{eqnarray}
t_\mu\,F^{({\rm III})}_\mu(t)\Big|_{t_\perp=0} & = &
i\,8\pi^2\,N_C\;\bar\sigma \sum_{f=u,d}\frac{1}{B_f^2}
\,\int_{q_\parallel\,q_\perp\,p_\perp\,p'_\perp}
\Big[\, g({s_\perp^2 + q_\parallel^+}^2) + g(s_\perp^2 + {q_\parallel^-}^2) -
\nonumber \\
& & 2\, g(s_\perp^2 + q_\parallel^2)\,\Big]\;g(q^2)\;
\exp\big[-\!i2\phi(q_\perp,p_\perp,p'_\perp)/(q_f B)\big] \,\times
\nonumber \\
& & \trmin_D \Big[\tilde S_f(p_\perp,q_\parallel^+)\,i\gamma_5\,
\tilde S_f(p'_\perp,q_\parallel^-)\,i\gamma_5\Big]\ ,
\end{eqnarray}
where $\phi(q_\perp,p_\perp,p'_\perp)$ is given by Eq.~(\ref{phi}), and we
have defined $s_\perp = p'_\perp + p_\perp - q_\perp$. Comparing with
Eq.~(\ref{f0k}), it is seen that the calculation to be done is basically the
same as that carried out in the case of the analysis of the $\pi^0$ mass,
described in Appendix A. In this way we obtain
\begin{eqnarray}
t_\mu\,F^{({\rm III})}_\mu(t)\Big|_{t_\perp=0} & = &
-\,i\,\frac{N_C}{2\pi}\;\bar\sigma \sum_{f=u,d}B_f
\sum_{k=0}^\infty\; \int_{q_\parallel}
\; \bigg[ \sum_{\lambda = \pm}\, g_{k,q_\parallel}^{\lambda,f}\,
\tilde g_{k,q_\parallel}^{\lambda,f}\,
F_{kk,q_\parallel^+q_\parallel^-}^{\lambda,f\,(AB)} +
\nonumber \\
& & 2\,k\,B_f\,\big(
g_{k,q_\parallel}^{+,f}\,\tilde g_{k,q_\parallel t_\parallel}^{-,f} +
g_{k,q_\parallel}^{-,f}\,\tilde g_{k,q_\parallel t_\parallel}^{+,f}\,\big)\,
F_{kk,q_\parallel^+q_\parallel^-}^{+,f\,(CD)}\bigg]
\ ,
\label{f3}
\end{eqnarray}
where we have defined
\begin{equation}
\tilde g_{k,q_\parallel t_\parallel}^{\lambda,f} \ = \ g_{k,q_\parallel^+}^{\lambda,f} +
g_{k,q_\parallel^-}^{\lambda,f} - 2\,g_{k,q_\parallel}^{\lambda,f}\ .
\end{equation}
Notice that one has
\begin{equation}
\bar\sigma\,\tilde g_{k,q_\parallel t_\parallel}^{\lambda,f} \ = \
M_{k,q_\parallel^+}^{\lambda,f} +
M_{k,q_\parallel^-}^{\lambda,f} - 2\,
M_{k,q_\parallel}^{\lambda,f}\ .
\end{equation}

When summing the contributions given by Eqs.~(\ref{f1}), (\ref{f2}) and
(\ref{f3}) it is seen that some cancellations help to simplify the final
expression for $t\cdot F(t)|_{t_\perp = 0}\,$. After some algebra one gets
\begin{eqnarray}
t_\mu\,F_\mu(t)\Big|_{t_\perp=0} & = &
i\,\frac{N_C}{\pi}\;\sum_{f=u,d}B_f
\sum_{k=0}^\infty\; \int_{q_\parallel}
\; \bigg[ \sum_{\lambda = \pm}\, g_{k,q_\parallel}^{\lambda,f}
\Big(F_{kk,q_\parallel^+q_\parallel^-}^{\lambda,f\,(AB)}\,
M_{k,q_\parallel}^{\lambda,f} - \hat A_{k,q_\parallel}^{\lambda,f}\Big) +
\nonumber \\
& & 2\,k\,B_f\,\Big(\,
g_{k,q_\parallel}^{+,f}\,M_{k,q_\parallel}^{-,f} +
g_{k,q_\parallel}^{-,f}\,M_{k,q_\parallel}^{+,f}\,\Big)\,
F_{kk,q_\parallel^+q_\parallel^-}^{+,f\,(CD)}\bigg]
\ .
\label{fsuma}
\end{eqnarray}
Moreover, the expression for $f_{\pi^0}$ can be further simplified by making use
of the gap equation and the relation (\ref{pimass}) obtained for the $\pi^0$
mass. According to the result previously obtained in
Ref.~\cite{GomezDumm:2017iex}, the gap equation can be written as
\begin{equation}
\frac{\bar \sigma}{G} \ = \ \frac{N_C}{\pi}\sum_{f=u,d} B_f
\sum_{k=0}^\infty\; \int_{q_\parallel}\,
\sum_{\lambda = \pm}\, g_{k,q_\parallel}^{\lambda,f}
\, \hat A_{k,q_\parallel}^{\lambda,f}\;\ ,
\label{gapeq}
\end{equation}
while for the pion mass we have
\begin{equation}
\frac{1}{G} \ = \ - F(0,-m_{\pi^0}^2) \ ,
\label{pionmass}
\end{equation}
with $F(0,t_\parallel^2)$ given by Eq.~(\ref{f0kfinal}). Taking into account
these equations and the relation in Eq.~(\ref{mmain}), it is easy to see
that for $t_\parallel^2=-m_{\pi^0}^2$ there are some additional cancellations in
Eq.~(\ref{fsuma}). Thus, we arrive to our final expression
\begin{equation}
m_{\pi^0}^2 \, f_{\pi^0} \ = \ m_c\,Z_{\pi^0}^{1/2}\,J(-m_{\pi^0}^2)\ ,
\label{ffinal}
\end{equation}
where the function $J(t_\parallel^2)$ is given by
\begin{equation}
J(t_\parallel^2) \ = \ \frac{N_C}{\pi} \sum_{f=u,d}
B_f\sum_{m=0}^\infty \int_{q_\parallel} \sum_{\lambda = \pm}\,
g^{\lambda,f}_{k,q_\parallel}\Big(
F_{kk,q_\parallel^+q_\parallel^-}^{\lambda,f\,(AB)} +
2k\,B_f\,F_{kk,q_\parallel^+q_\parallel^-}^{\lambda,f\,(CD)} \Big)\ ,
\label{fj}
\end{equation}
with $q_\parallel^\pm = q_\parallel\pm t_\parallel$. Taking the limit $B\to 0$
one arrives at the expression given e.g.\ in Ref.~\cite{GomezDumm:2006vz},
\begin{equation}
J(t^2)\bigg|_{B=0} \ = \ 8N_C\,\int\ \frac{d^4q}{(2\pi)^4}\ g(q^2)\,\frac{(q^+\cdot
q^-)\, +\, M({q^+}^2)\,M({q^-}^2)}{\left[{q^+}^2 + M({q^+}^2)\right]
\left[{q^-}^2 + M({q^-}^2)\right]}\ .
\label{jb0}
\end{equation}

\subsection{Chiral relations}

In this subsection we show that the Goldberger-Treiman (GT) and
Gell-Mann-Oakes-Renner (GOR) relations remain valid in our model in the
presence of the external magnetic field. For this purpose, following the
line of the analysis in Ref.~\cite{GomezDumm:2006vz}, it is useful to define
the function
\begin{equation}
K(t_\parallel^2) \ = \ m_c\,J(t_\parallel^2)-\bar\sigma F(0,t_\parallel^2)\ ,
\end{equation}
where $J(t_\parallel^2)$ and $F(0,t_\parallel^2)$ are given by
Eqs.~(\ref{fsuma}) and (\ref{f0kfinal}), respectively. From
Eq.~(\ref{fsuma}), taking into account the relation in Eq.~(\ref{mmain}) it
is easy to show that
\begin{equation}
-i\,t_\mu\,F_\mu(t)\Big|_{t_\perp=0} \ = \
K(t_\parallel^2) - \frac{N_C}{\pi}
\sum_{f=u,d}B_f
\sum_{k=0}^\infty\; \int_{q_\parallel}
\sum_{\lambda = \pm}\, g_{k,q_\parallel}^{\lambda,f}
\,\hat A_{k,q_\parallel}^{\lambda,f}\ .
\end{equation}
The second term in the rhs is a constant, equal to $-\bar \sigma/G$ according
to the gap equation. Moreover, taking into account the relations
\begin{eqnarray}
F_{kk,q_\parallel q_\parallel}^{\lambda,f\,(AB)} +
2\,k\,B_f\,F_{kk,q_\parallel q_\parallel}^{\lambda,f\,(CD)} & = &
\hat B_{k,q_\parallel}^{\lambda,f}\ , \nonumber \\
\big(M_{k,q_\parallel}^{\pm,f} - M_{k,q_\parallel}^{\mp,f}\big)
\,F_{kk,q_\parallel q_\parallel}^{\lambda,f\,(CD)} & = & \hat
D_{k,q_\parallel}^{\pm,f}\ , \nonumber \\
\hat B_{k,q_\parallel}^{\lambda,f} M_{k,q_\parallel}^{\lambda,f}
- 2k\,B_f\,\hat D_{k,q_\parallel}^{\lambda,f} & = &
\hat A_{k,q_\parallel}^{\lambda,f}\ ,
\label{varios}
\end{eqnarray}
it is seen that
\begin{equation}
m_c\,J(0)-\bar\sigma F(0,0) \ = \ \frac{N_C}{\pi}
\sum_{f=u,d}B_f
\sum_{k=0}^\infty\; \int_{q_\parallel}
\sum_{\lambda = \pm}\, g_{k,q_\parallel}^{\lambda,f}
\,\hat A_{k,q_\parallel}^{\lambda,f}\ ,
\end{equation}
hence we can write
\begin{equation}
-i\,t_\mu\,F_\mu(t)\Big|_{t_\perp=0} \ = \
K(t_\parallel^2) - K(0)\ .
\end{equation}
Thus, from Eq.~(\ref{fpi}) we obtain
\begin{equation}
f_{\pi^0} \ = \ -\,Z_{\pi^0}^{1/2}\,\frac{\big[ K(-m_{\pi^0}^2) - K(0) \big]}
{-m_{\pi^0}^2}\ .
\end{equation}

In the chiral limit one has $m_c\to 0$, $m_\pi^2\to 0$, therefore the
pion decay constant is given by
\begin{equation}
f_{\pi^0,0} \ = \ -\,Z_{\pi^0,0}^{1/2}\, \frac{dK_0(t_\parallel^2)}{dt_\parallel^2}\bigg|_{t_\parallel = 0}
\ = \ Z_{\pi^0,0}^{1/2}\,\bar\sigma_0\,
\frac{dF_0(0,t_\parallel^2)}{dt_\parallel^2}\bigg|_{t_\parallel = 0}\ = \
Z_{\pi^0,0}^{-1/2}\,\bar\sigma_0\ ,
\label{fpicero}
\end{equation}
where we have taken into account the relation between $Z_{\pi^0}$ and the
derivative of $F(0,t_\parallel^2)$ in Eq.~(\ref{zpi}). Subindices 0 indicate
that all quantities have to be evaluated in the chiral limit. Noticing
that $Z_{\pi^0}^{1/2}$ turns out to be the effective coupling constant $g_{\pi
q\bar q}$ between the $\pi_3$ field and the quark-antiquark pseudoscalar
currents, we arrive at
\begin{equation}
f_{\pi^0,0}\, g_{\pi q\bar q,0}\ = \ \bar\sigma_0\ ,
\end{equation}
which is the expression for the Goldberger-Treiman relation at the quark
level.

Finally, let us consider the quark condensates, $\langle \bar uu\rangle$ and
$\langle \bar dd\rangle$, which in the presence of the magnetic field are
given by Eq.~(\ref{conden}). Taking into account the relations
(\ref{varios}), it is easy to see that in the chiral limit one has
\begin{equation}
\langle \bar uu + \bar dd\rangle_0 \ = \ -\,\bar\sigma_0\, J_0(0)
\end{equation}
[notice that away from the chiral limit the integrals in Eq.~(\ref{conden})
are in general divergent, and need to be regularized]. In addition, we can
perform a chiral expansion at both sides of Eq.~(\ref{ffinal}), keeping only
the lowest nonzero order. This leads to
\begin{equation}
m_{\pi^0}^2 \, f_{\pi^0,0} \ = \ m_c\,Z_{\pi^0,0}^{1/2}\,J_0(0)\ .
\end{equation}
{}From this relation, together with Eq.~(\ref{fpicero}), we obtain the
Gell-Mann-Oakes-Renner relation for the $\pi^0$ meson,
\begin{equation}
m_c\,\langle \bar uu + \bar dd\rangle_0 \ = \ -\,m_{\pi^0}^2 \, f_{\pi^0,0}^2 \ .
\label{gor}
\end{equation}

\section{Numerical results}

To obtain definite numerical predictions for the behavior of the above
defined quantities as functions of the external magnetic field, it is
necessary to specify the particular shape of the nonlocal form factor
$g(p^2)$. We consider here two often-used
forms~\cite{Scarpettini:2003fj,GomezDumm:2006vz,GomezDumm:2001fz}, namely a
Gaussian function
\begin{equation}
g(p^2) \ = \ \exp(-p^2/\Lambda^2)
\end{equation}
and a ``5-Lorentzian'' function
\begin{equation}
g(p^2) \ = \ \frac{1}{1 + (p^2/\Lambda^2)^5}\ \ .
\end{equation}
Notice that in the form factors we introduce an energy scale $\Lambda$,
which acts as an effective momentum cut-off. This has to be taken as a free
parameter of the model, together with the current quark mass $m_c$ and the
coupling constant $G$ in the effective Lagrangian. In the particular case of
the Gaussian form factor one has the advantage that the integral in
Eq.~(\ref{mpk}) can be performed analytically, allowing to a dramatic
reduction of the computer time needed for numerical calculations of
the relevant quantities.

As in Refs.~\cite{Pagura:2016pwr,GomezDumm:2017iex} (see also the discussion
on different parameterizations in Ref.~\cite{GomezDumm:2006vz}), we
determine the free parameters by requiring the model to reproduce the
empirical values of the pion mass and decay constant, as well as some
phenomenologically adequate value of the quark condensate $\langle \bar
ff\rangle_{\rm reg}$, at $B = 0$ [the pion mass and decay constant in the
limit $B=0$ can be calculated from Eqs.~(\ref{fb0}) and (\ref{jb0})]. The
parameter sets obtained for Gaussian and 5-Lorentzian form factors,
considering different values of the condensate, can be found in
Ref.~\cite{GomezDumm:2017iex}. In that article, the behavior of the chiral
quark condensates with the magnetic field has been analyzed, showing that at
zero temperature the condensates grow monotonically with $B$ (magnetic
catalysis). Moreover, it is seen that these curves turn out to be in good
quantitative agreement with the results obtained from LQCD calculations. The
agreement is found to be particularly accurate for the parameter sets $m_c =
6.5$~MeV, $\Lambda = 678$~MeV, $G\Lambda^2 = 23.66$ and $m_c = 6.5$~MeV,
$\Lambda = 857$~MeV, $G\Lambda^2 = 9.700$, corresponding to $\langle \bar
ff\rangle_{\rm reg} = (- 230~\rm{MeV})^3$ for Gaussian and 5-Lorentzian form
factors, respectively.

\begin{figure}[hbt]
\includegraphics[width=0.7\textwidth]{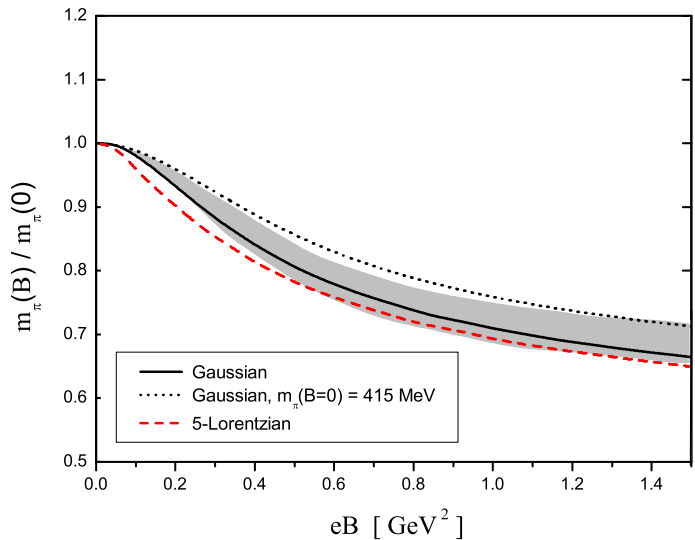}
\caption{Mass of the $\pi^0$ meson as a function of $eB$, normalized to its
value for $B=0$. Solid and dashed lines correspond to Gaussian and
5-Lorentzian form factors, respectively. The dotted line is obtained for a
parameterization in which $m_\pi = 415$~MeV, while the gray band corresponds
to the results of lattice QCD calculations quoted in
Ref.~\cite{Bali:2017ian}.} \label{fig1}
\end{figure}

\begin{figure}[hbt]
\includegraphics[width=0.7\textwidth]{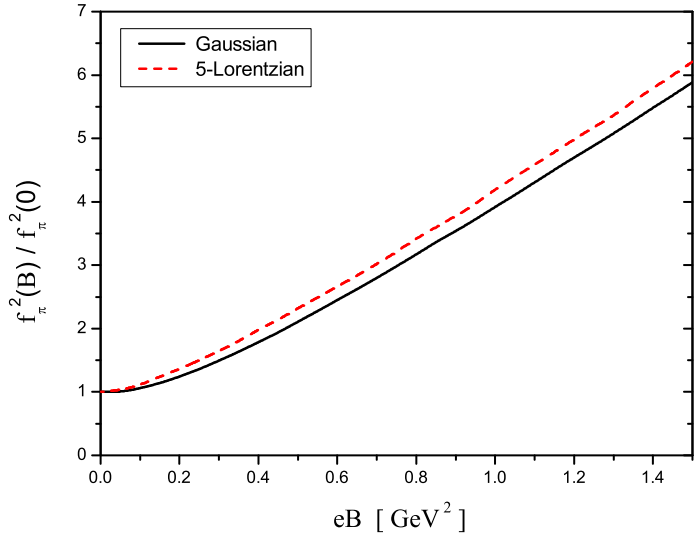}
\caption{Normalized squared pion decay coupling $f_{\pi^0}^2$ as a function
of the external magnetic field, for Gaussian and 5-Lorentzian form factors.}
\label{fig2}
\end{figure}

Our results for the behavior of the pion mass $m_{\pi^0}(B)$ and the squared
pion decay constant $f_{\pi^0}^{\,2}(B)$ for the above mentioned parameter
sets are shown in Figs.~\ref{fig1} and \ref{fig2}, respectively. In both
cases the curves have been normalized to $B=0$ values $m_{\pi^0}(0)= 139$
MeV and $f_{\pi^0}^{\,2} = (92.4\ {\rm MeV})^2$. As shown in
Fig.~\ref{fig1}, the $\pi^0$ mass is found to decrease when $eB$ gets
increased, reaching a value of about 65\% of $m_{\pi^0}(0)$ at $eB\simeq
1.5$ GeV$^2$, which corresponds to a magnetic field of about $2.5\times
10^{20}$~G. We also include in Fig.~\ref{fig1} a gray band that corresponds
to recently quoted results from lattice QCD~\cite{Bali:2017ian}. The latter
have been obtained from a continuum extrapolation of lattice spacing,
considering a relatively large quark mass for which $m_\pi = 415$~MeV. For
comparison, we also quote the results obtained within our model by shifting
$m_c$ to 56.3~MeV, which leads to this enhanced pion mass. In general it is
seen from the figure that our predictions turn out to be in good agreement
with LQCD calculations. It is worth remarking that our results have been
obtained directly from model parameterizations used in previous works (where
external magnetic fields have not been taken into
account)~\cite{GomezDumm:2006vz}, i.e.~no extra adjustments have been
performed to fit LQCD data. This is in contrast to the situation in the
local NJL model, in which comparable results for the pion mass behavior are
obtained after introducing a $B$ dependent coupling constant adjusted to
reproduce LQCD results for the quark condensates~\cite{Avancini:2016fgq}.
Concerning the pion decay constant $f_{\pi^0}$, as shown in Fig.~\ref{fig2}
we find that it behaves as an increasing function of $B$. This is fully
consistent with the approximate validity of the Gell-Mann-Oakes-Renner
relation for a small value of the constituent mass $m_c$. In fact, taking
into account the behavior of the $\pi^0$ mass, from Eq.~(\ref{gor}) it is
seen that $f_{\pi^0}^2$ should grow somewhat more rapidly than the
condensates, which is in agreement with the results in Fig.~\ref{fig2} (the
curves showing the behavior of the condensates can be found in
Ref.~\cite{GomezDumm:2017iex}). For example, at $eB = 1.5$~GeV$^2$ one gets
$m_c\langle \bar uu+\bar dd\rangle /(m_{\pi^0}^2f_{\pi^0}^2) \simeq - 0.98$,
both for Gaussian and 5-Lorentzian form factors. It is also worth mentioning
that the curves in Figs.~\ref{fig1} and \ref{fig2} are found to remain
practically unchanged when the value of the $B=0$ condensate used to fix the
parameterization is varied within the range from $-(220~{\rm MeV})^3$ to
$-(250~{\rm MeV})^3$.

\begin{figure}[hbt]
\includegraphics[width=0.7\textwidth]{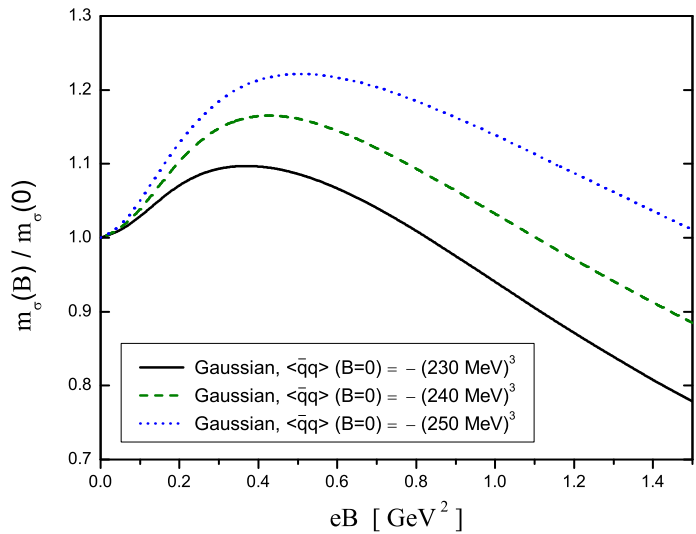}
\caption{Mass of the $\sigma$ meson as a function of $eB$, normalized to its
value for $B=0$, for three different parameterizations (all of them
corresponding to a Gaussian form factor).} \label{fig3}
\end{figure}

Finally, in Fig.~\ref{fig3} we quote the values of the sigma meson mass for
$eB$ up to 1.5~GeV$^2$, normalized to $m_\sigma(0)$. In the case of the
sigma mass the results turn out to be more dependent on the parameter set,
therefore we consider here three different parameterizations leading to
$\langle \bar ff\rangle_{\rm reg}|_{(B=0)}^{1/3} = -230$, $-240$ and
$-250$~MeV, for the Gaussian form factor. The corresponding values of
$m_\sigma$ for $B=0$ are 771, 683 and 616 MeV, respectively. For lower
values of the $B=0$ condensates, as well as for the case of 5-Lorentzian
form factors, the determination of the $\sigma$ mass becomes problematic
since it exceeds a threshold of formation of two on-shell quarks, which
requires an additional regularization prescription. This problem is usually
found in NJL-like theories when one deals with relatively large meson
masses. From Fig.~\ref{fig3} we observe that for all the cases considered
the $\sigma$ meson mass shows a nonmonotonic behavior as a function of $B$.
Namely, it gets increased for low $B$, reaching a maximum at about $eB=0.4 \
\mbox{GeV}^2$, after which it shows a steady decrease. It is worth noticing
that a qualitative similar behavior is obtained within the local NJL model
when a $B$ dependent coupling constant is introduced~\cite{Avancini:2016fgq}.

\section{Summary and conclusions}

We have studied the behavior of neutral meson properties in the presence of
a uniform static external magnetic field $B$ in the context of a nonlocal
chiral quark model. In this approach, which can be viewed as an extension of
the local Nambu-Jona-Lasinio model, the effective couplings between
quark-antiquark currents include nonlocal form factors that regularize
ultraviolet divergences in quark loop integrals and lead to a
momentum-dependent effective mass in quark propagators. We have worked out
the formalism introducing Ritus transforms of Dirac fields, which allow to
obtain closed analytical expressions for meson polarization functions and
for the pion decay constant. In addition, we have shown that the
Goldberger-Treiman and Gell-Mann-Oakes-Renner chiral relations remain valid
within this model in the presence of the external magnetic field. In our
numerical calculations we have considered the case of Gaussian and
Lorentzian form factors, choosing some sets of model parameters that allow
to reproduce the empirical values of the pion mass and decay constants and
lead to acceptable values of the quark condensate for $B=0$. Our results for
the neutral pion mass behavior with the magnetic field display a very mild
dependence on the parametrization and/or form factor and turn out to be in
good quantitative agreement with the available lattice QCD calculations. In
the case of the pion decay constant, our results are also quite independent
of the chosen parametrization, displaying a rather strong increase of
$f_{\pi^0}$ with $eB$ that implies, for example, $f_{\pi^0}(1\mbox{~GeV}^2)
\simeq 2 \ f_{\pi^0}(0)$. On the other hand, our results for the sigma mass
behavior with the magnetic field show a stronger dependence on the
parametrization. Nonetheless, in all the cases considered it is seen that
$m_\sigma$ shows a nonmonotonic behavior as a function of $B$. A qualitative
similar behavior is obtained within the local NJL model when a $B$-dependent
coupling constant is introduced~\cite{Avancini:2016fgq}.

We conclude by noting that, given the present results for the neutral pion
mass and the fact that nonlocal chiral quark models naturally lead to the
Inverse Magnetic Catalysis effect~\cite{Pagura:2016pwr,GomezDumm:2017iex},
an extension of the present work to finite temperature appears to be very
interesting. The study of the behavior of the charged pion properties within
the present framework, although more involved due to the corresponding
Schwinger phase structure, also deserves further attention. We expect to
report on these issues in forthcoming articles.

\section*{Acknowledgements}

This work has been supported in part by CONICET and ANPCyT (Argentina),
under grants PIP14-492, PIP12-449, and PICT14-03-0492, and by the National
University of La Plata (Argentina), Project No.\ X718.

\section*{Appendix A}

\newcounter{erasmo}
\renewcommand{\thesection}{\Alph{erasmo}}
\renewcommand{\theequation}{\Alph{erasmo}\arabic{equation}}
\setcounter{erasmo}{1} \setcounter{equation}{0} 

We outline here the derivation of the relation in Eq.~(\ref{f0kfinal}). It
is easy to see that the expression in Eq.~(\ref{f0k}) can be rearranged in
the form
\begin{equation}
F(0,k_\parallel^2) \ = \ -\,128\,\pi^2\,N_C \sum_{f=u,d}\,\frac{1}{B_f^2}
\sum_{k,k'=0}^\infty \int_{q_\parallel} \ \bigg[
\sum_{\lambda = \pm}\,F_{kk',q_\parallel^+ q_\parallel^-}^{\lambda,f\,(AB)}\,
I_{kk',q_\parallel}^{\lambda,f\,(0)} + F_{kk',q_\parallel^+ q_\parallel^-}^{+,f\,(CD)}\,
I_{kk',q_\parallel}^{f\,(1)}\bigg]
\ ,
\label{f0kaux}
\end{equation}
where
\begin{eqnarray}
\hspace{-0.7cm} I_{kk',q_\parallel}^{\lambda,f\,(0)} & = & (-1)^{k+k'}\int_{q_\perp\, p_\perp\, p'_\perp}
\,\exp[i2\phi(q_\perp,p_\perp,p'_\perp)/(q_fB)] \,\exp[-(p_\perp^2+{p'_\perp}^{\!\!
2})/B_f]\, \times \nonumber \\
& & g(q_\perp^2 + q_\parallel^2)\, g[(p'_\perp + p_\perp - q_\perp)^2\!+
q_\parallel^2]\,L_{k_\lambda}(2p_\perp^2/B_f)\, L_{k'_\lambda}(2{p'_\perp}^{\!\!2}/B_f)\ , \\
\hspace{-0.7cm} I_{kk',q_\parallel}^{f\,(1)} & = & 8\,(-1)^{k+k'} \int_{q_\perp\, p_\perp\, p'_\perp}
\,\exp[i2\phi(q_\perp,p_\perp,p'_\perp)/(q_fB)] \,\exp[-(p_\perp^2+{p'_\perp}^{\!\!
2})/B_f]\,\times \nonumber \\
& & (p_\perp\cdot p'_\perp)\,g(q_\perp^2 + q_\parallel^2)\, g[(p'_\perp + p_\perp - q_\perp)^2\!+
q_\parallel^2]\,L^1_{k-1}(2p_\perp^2/B_f)\,
L^1_{k'-1}(2{p'_\perp}^{\!\!2}/B_f)\ .
\label{imms}
\end{eqnarray}
These integrals can be worked out by taking the Laguerre-Fourier transforms
of the nonlocal form factors given by Eq.~(\ref{inversa}). We obtain in this way
\begin{eqnarray}
I_{kk',q_\parallel}^{\lambda,f\,(0)} & = & 4\,(-1)^{k+k'}\sum_{m,m'=0}^\infty
(-1)^{m+m'}\,g_{m,q_\parallel}^{\lambda,f}\,g_{m',q_\parallel}^{\lambda,f}\int_{q_\perp\, p_\perp\, p'_\perp}
\,\exp[i2\phi(q_\perp,p_\perp,p'_\perp)/(q_fB)]\,\times \nonumber \\
& & \exp[-(p_\perp^2+{p'_\perp}^{\!\! 2}+q_\perp^2+(p'_\perp+p_\perp-q_\perp)^2)/B_f]\, \times \nonumber\\
& & L_{k_\lambda}(2p_\perp^2/B_f)\, L_{k'_\lambda}(2{p'_\perp}^{\!\!2}/B_f)\, L_{m_\lambda}(2q_\perp^2/B_f)
\,L_{m'_\lambda}[2(p'_\perp+p_\perp-q_\perp)^2/B_f] \ ,\\
I_{kk',q_\parallel}^{f\,(1)} & = & 32\,(-1)^{k+k'}\sum_{m,m'=0}^\infty
(-1)^{m+m'}\,g_{m,q_\parallel}^{+,f}\,g_{m',q_\parallel}^{-,f}\int_{q_\perp\, p_\perp\, p'_\perp}
\,\exp[i2\phi(q_\perp,p_\perp,p'_\perp)/(q_fB)]\,\times \nonumber \\
& & \exp[-(p_\perp^2+{p'_\perp}^{\!\! 2}+q_\perp^2+(p'_\perp+p_\perp-q_\perp)^2)/B_f]\,(p_\perp\cdot p'_\perp)\,
\times \nonumber\\
& & L^1_{k-1}(2p_\perp^2/B_f)\, L^1_{k'-1}(2{p'_\perp}^{\!\!2}/B_f) \, L_{m_+}(2q_\perp^2/B_f)
\,L_{m'_-}[2(p'_\perp+p_\perp-q_\perp)^2/B_f]\ .
\end{eqnarray}
Let us now change the integration variables, defining dimensionless two
dimensional vectors $u = -\sqrt{(2/B_f)}\, p_\perp$, $v = \sqrt{(2/B_f)}\,
p'_\perp$, $w = \sqrt{(2/B_f)}\, (p_\perp - q_\perp)$. The integrals read
\begin{eqnarray}
I_{kk',q_\parallel}^{\lambda,f\,(0)} & = & \frac{B_f^3}{2}\,(-1)^{k+k'}\sum_{m,m'=0}^\infty
(-1)^{m+m'}\,g_{m,q_\parallel}^{\lambda,f}\,g_{m',q_\parallel}^{\lambda,f}\, K_{kk'mm'}^{\lambda,f\,(0)}\ ,
\nonumber \\
I_{kk',q_\parallel}^{f\,(1)} & = & 2\,B_f^4\,(-1)^{k+k'}\sum_{m,m'=0}^\infty
(-1)^{m+m'-1}\,g_{m,q_\parallel}^{+,f}\,g_{m',q_\parallel}^{-,f}\, K_{kk'mm'}^{f\,(1)}\ ,
\end{eqnarray}
where
\begin{eqnarray}
K_{kk'mm'}^{\lambda,f\,(0)} & = & \int_{u\, v\, w} \!\!\!\exp[-w^2]\,\exp[-u^2\!-u\cdot w - is_f(u_1w_2-u_2w_1)]\,
L_{k_\lambda}(u^2)\,L_{m_\lambda}[(u+w)^2]\times \nonumber \\
& & \qquad\exp[-v^2\!-v\cdot w -
is_f(v_1w_2-v_2w_1)]\,L_{k'_\lambda}(v^2)L_{m'_\lambda}[(v+w)^2]\ ,
\nonumber\\
K_{kk'mm'}^{f\,(1)} & = & - \!\!\int_{u\, v\, w} \!\!\!\exp[-w^2]\,\exp[-u^2\!-u\cdot w - is_f(u_1w_2-u_2w_1)]\,
L_{k-1}^1(u^2)\,L_{m_+}[(u+w)^2]\times \nonumber \\
& & \qquad (u\cdot v) \, \exp[-v^2\!-v\cdot w - is_f(v_1w_2-v_2w_1)]\,L_{k'-1}(v^2)L_{m'_-}[(v+w)^2]\ .
\end{eqnarray}
Notice that $K_{kk'mm'}^{\lambda,f\,(0)}$ and $K_{kk'mm'}^{f\,(1)}$ do not
depend on momenta, nor on the magnetic field. Their calculation can be
performed with the aid of the following useful relations,
\begin{eqnarray}
\!\!\!
\frac{1}{2\pi}\int_0^{2\pi}\!\! d\theta\; L_n(x^2\!+y^2\!+2 xy \cos\theta) \,\exp[-xy\exp(\pm i\theta)]
& = & L_n(x^2)\,L_n(y^2)\ ,
\label{master} \\
\!\!\!\frac{1}{2\pi}\int_0^{2\pi}\!\! d\theta\;\cos\theta\, L_n(x^2\!+y^2\!+2 xy \cos\theta)
\,\exp[-xy\exp(\pm i\theta)]
& = & -\,\frac{xy}{2}\Big[\frac{L_n^1(x^2)\,L_n^1(y^2)}{n+1}+\nonumber \\
& & \frac{L_{n-1}^1(x^2)\,L_{n-1}^1(y^2)}{n}\Big]\ ,
\label{master2} \\
\!\!\!\frac{1}{2\pi}\int_0^{2\pi}\!\! d\theta\;\sin\theta\, L_n(x^2\!+y^2\!+2 xy \cos\theta)
\,\exp[-xy\exp(\pm i\theta)]
& = & \mp\,\frac{ixy}{2}\Big[\frac{L_n^1(x^2)\,L_n^1(y^2)}{n+1}-\nonumber \\
& & \frac{L_{n-1}^1(x^2)\,L_{n-1}^1(y^2)}{n}\Big]\ ,
\label{master3}
\end{eqnarray}
together with the orthogonality properties of the generalized Laguerre
polynomials. In the case of $K_{kk'mm'}^{\lambda,f\,(0)}$, usage of
Eq.~(\ref{master}) leads to
\begin{eqnarray}
K_{kk'mm'}^{\lambda,f\,(0)} & = & \frac{1}{(4\pi)^2}\int_{w} \exp(-w^2)\int_0^\infty du^2
\;\exp(-u^2)\,L_{k_\lambda}(u^2)\,L_{m_\lambda}(u^2)\,L_{m_\lambda}(w^2)\times \nonumber \\
& & \int_0^\infty dv^2
\;\exp(-v^2)\,L_{k'_\lambda}(v^2)\,L_{m'_\lambda}(v^2)\,L_{m'_\lambda}(w^2)\nonumber \\
& = & \frac{1}{(4\pi)^3}\;\delta_{km}\,\delta_{k'm'}\,\delta_{mm'}\ ,
\end{eqnarray}
and consequently
\begin{equation}
I_{kk',q_\parallel}^{\lambda,f\,(0)} \ = \ \frac{B_f^3}{128\,\pi^3}\;g_{k,q_\parallel}^{\lambda,f}
\,g_{k,q_\parallel}^{\lambda,f}\,\delta_{kk'}\ .
\label{i0}
\end{equation}
Finally, using Eqs.~(\ref{master2}) and (\ref{master3}) we obtain
\begin{equation}
K_{kk'mm'}^{f\,(1)} \ = \ -\,\frac{1}{128\pi^3}\,k\,\delta_{kk'}\,
\big(\delta_{m+1\, k_-}\,\delta_{m'k_+} + \delta_{m
k_-}\,\delta_{m'-1\,k_+}\big)\ ,
\end{equation}
which leads to
\begin{equation}
I_{kk',q_\parallel}^{f\,(1)} \ = \ \frac{k\, B_f^4}{32\pi^3}\;g_{k,q_\parallel}^{+,f}
\,g_{k,q_\parallel}^{-,f}\,\delta_{kk'}\ .
\label{i1}
\end{equation}
Replacing the results in Eq.~(\ref{i0}) and (\ref{i1}) in Eq.~(\ref{f0kaux})
one arrives at our final expression, quoted in Eq.~(\ref{f0kfinal}).

\section*{Appendix B}

\newcounter{erasmo2}
\renewcommand{\thesection}{\Alph{erasmo2}}
\renewcommand{\theequation}{\Alph{erasmo2}\arabic{equation}}
\setcounter{erasmo2}{2} \setcounter{equation}{0} 

Let us discuss here the derivation of our results in Eqs.~(\ref{f2}) and
(\ref{f3}). We start from the expression in Eq.~(\ref{f2ini}). Introducing
the Laguerre-Fourier transform of $g(q^2)$ and changing the order of
integrals and sums one gets
\begin{eqnarray}
\hspace{-1cm}t_\mu\,F^{({\rm II})}_\mu(t)\Big|_{t_\perp=0} & = &
i\,128\,\pi^2N_C\sum_{f=u,d}\frac{1}{B_f^2}\sum_{k,k',m=0}^\infty
\int_{q_\parallel}\bigg\{\sum_{\lambda = \pm}\;
g_{m,q_\parallel}^{\lambda,f}\,\times
\nonumber \\
& & \Big[(t_\parallel \cdot q_\parallel^-)\,\hat A_{k,q_\parallel^+}^{\lambda,f}
\,\hat B_{k',q_\parallel^-}^{\lambda,f} -
(t_\parallel \cdot q_\parallel^+)\,\hat A_{k',q_\parallel^-}^{\lambda,f}
\,\hat B_{k,q_\parallel^+}^{\lambda,f}\Big]\,
\tilde K_{kk'm}^{\lambda,f\,(0)} \ + \nonumber \\
& & 8\,i\,g_{m,q_\parallel}^{+,f}\,\Big[ (t_\parallel \cdot q_\parallel^-)
\,\hat C_{k,q_\parallel^+}^{+,f} \,\hat D_{k',q_\parallel^-}^{+,f} -
(t_\parallel \cdot q_\parallel^+)\,\hat C_{k',q_\parallel^-}^{+,f}
\,\hat D_{k,q_\parallel^+}^{+,f}\Big]\,\tilde K_{kk'm}^{f\,(1)}
\bigg\}\ ,
\label{b1}
\end{eqnarray}
where
\begin{eqnarray}
\!\!\!\tilde K_{kk'm}^{\lambda,f\,(0)} & = & (-1)^{k+k'+m_\lambda}
\int_{q_\perp\, p_\perp\, p'_\perp}
\,\exp[-i2\phi(q_\perp,p_\perp,p'_\perp)/(q_fB)] \, \times \nonumber \\
& & \exp[-(p_\perp^2+{p'_\perp}^{\!\! 2}+q_\perp^2)/B_f]\,
L_{k_\lambda}(2p_\perp^2/B_f)\, L_{k'_\lambda}(2{p'_\perp}^{\!\!2}/B_f)\,L_{m_\lambda}(2q_\perp^2/B_f)\ , \\
\!\!\!\tilde K_{kk'm}^{f\,(1)} & = & (-1)^{k+k'+m_+}
\int_{q_\perp\, p_\perp\, p'_\perp}
\,\exp[-i2\phi(q_\perp,p_\perp,p'_\perp)/(q_fB)] \, (p_1p'_2-p_2p'_1)\,\times \nonumber \\
& & \exp[-(p_\perp^2+{p'_\perp}^{\!\! 2}+q_\perp^2)/B_f]\,
\,L^1_{k-1}(2p_\perp^2/B_f)\, L^1_{k'-1}(2{p'_\perp}^{\!\!2}/B_f)\,L_{m_+}(2q_\perp^2/B_f)\ .
\end{eqnarray}
Now we change the integration variables, defining dimensionless two
dimensional vectors $u = \sqrt{(2/B_f)}\, q_\perp$, $v = \sqrt{(2/B_f)}\,
p_\perp$, $w = \sqrt{(2/B_f)}\, (p'_\perp - p_\perp)$. The integrals read
\begin{eqnarray}
\!\!\!\tilde K_{kk'm}^{\lambda,f\,(0)} & = &
(-1)^{k+k'+m_\lambda}\,\frac{B_f^3}{8}
\int_{v\, w}
\,\exp[is_f(v_1w_2-v_2w_1)] \, \exp[-(v^2+v\cdot w+w^2/2)]\,\times \nonumber \\
& & L_{k_\lambda}(v^2)\,L_{k'_\lambda}[(v+w)^2]\,\int_u
\exp(-u^2/2)\, L_{m_\lambda}(u^2)\,\exp[is_f(w_1u_2-w_2u_1)]\ , \\
\!\!\!\tilde K_{kk'm}^{f\,(1)} & = & (-1)^{k+k'+m_+} \frac{B_f^4}{16}
\int_{v\, w}\,\exp[is_f(v_1w_2-v_2w_1)] \, \exp[-(v^2+v\cdot w+w^2/2)]\,\times \nonumber \\
& & (v_1w_2-v_2w_1)\, L_{k-1}^1(v^2)\,L_{k'-1}^1[(v+w)^2]\,\times \nonumber \\
& & \int_u \exp(-u^2/2)\, L_{m_+}(u^2)\,\exp[is_f(w_1u_2-w_2u_1)]\ .
\end{eqnarray}
To evaluate the integrals over $u$, let us fix the external vector $w$ along
the 1 direction. We get
\begin{eqnarray}
\int_u
\exp(-u^2/2)\, L_{m_\lambda}(u^2)\,\exp[is_f(w_1u_2-w_2u_1)] & &
\nonumber \\
& & \hspace{-7cm} = \
\frac{1}{(2\pi)^2}\int_0^\infty d|u|\,|u|\, \exp(-u^2/2)\, L_{m_\lambda}(u^2)
\int_0^{2\pi} d\theta\; \exp(is_f |w u|\sin\theta) \nonumber \\
& & \hspace{-7cm} = \
\frac{1}{2\pi}\int_0^\infty d|u|\,|u|\, \exp(-u^2/2)\, L_{m_\lambda}(u^2)\,
J_0(|w u|) \nonumber \\
& & \hspace{-7cm} = \ \frac{(-1)^{m_\lambda}}{2\pi}\,\exp(-w^2/2)\,
L_{m_\lambda}(w^2)\ ,
\label{intu}
\end{eqnarray}
where we have used the relations
\begin{equation}
\int_0^{2\pi} d\theta\; \exp(\pm i y \sin\theta) \ = \ 2\pi\,J_0(y)
\end{equation}
and
\begin{equation}
\int_0^\infty dx\;x^{\nu+1}\,e^{-\beta x^2}\,L_n^\nu(\alpha
x^2)\,J_\nu(xy) \ = \ \frac{(1-\alpha/\beta)^n}{(2\beta)^{\nu+1}}\;y^\nu\,
e^{-y^2/(4\beta)}\,L_n^\nu\Big[\frac{\alpha
y^2}{4\beta(\alpha-\beta)}\Big]\ ,
\end{equation}
$J_\nu(x)$ being Bessel functions of the first kind. Now, taking into
account Eq.~(\ref{master}), together with the orthogonality property of the
Laguerre polynomials, we find
\begin{eqnarray}
\!\!\!\tilde K_{kk'm}^{\lambda,f\,(0)} & = &
(-1)^{k+k'}\,\frac{B_f^3}{128\pi^4}
\int_0^\infty\, d|w|\,|w|\,\exp(-w^2)\,L_{m_\lambda}(w^2)
\,\int_0^\infty \,d|v|\,|v|\,\exp(-v^2)\, \times \nonumber \\
& & L_{k_\lambda}(v^2)\,\int_0^{2\pi} d\psi\, L_{k'_\lambda}(v^2+w^2+2\,|vw|\cos\psi)
\,\exp[-|vw|\exp(is_f\psi)] \nonumber \\
& = & (-1)^{k+k'}\,\frac{B_f^3}{64\pi^3}
\int_0^\infty\, d|w|\,|w|\,\exp(-w^2)\,L_{m_\lambda}(w^2)
L_{k'_\lambda}(w^2) \,\times \nonumber \\
& & \int_0^\infty \,d|v|\,|v|\,\exp(-v^2)\, L_{k_\lambda}(v^2)\, L_{k'_\lambda}(v^2) \nonumber \\
& = & \frac{B_f^3}{256\pi^3}\;\delta_{kk'}\,\delta_{k'm}\ .
\label{bk0}
\end{eqnarray}

For the evaluation of $\tilde K_{kk'm}^{f\,(1)}$ we use the result in
Eq.~(\ref{intu}) and then change to new variables $\bar v = -v$ and $\bar w = w+v$.
We have
\begin{eqnarray}
\!\!\!\tilde K_{kk'm}^{f\,(1)} & = &
(-1)^{k+k'}\,\frac{B_f^4}{256\,\pi^4}
\int_0^\infty\, d|\bar w|\,\bar w^2\,\exp(-\bar w^2)\,L_{k'-1}^1(\bar w^2)
\,\int_0^\infty \,d|\bar v|\,\bar v^2\,\exp(-\bar v^2)\, \times \nonumber \\
& & L_{k-1}^1(\bar v^2)\,\int_0^{2\pi} d\psi\,\sin\psi\,
L_{m_+}(\bar v^2+\bar w^2+2\, |\bar v \bar w|\cos\psi)
\,\exp[-|\bar v \bar w|\exp(-is_f\psi)] \nonumber \\
& = & (-1)^{k+k'}\,is_f\,\frac{B_f^4}{256\,\pi^3}
\bigg[ \frac{1}{m_++1}\int_0^\infty\, d|\bar w|\,|\bar w|^3\,\exp(-\bar w^2)\,L_{k'-1}^1(\bar w^2)
L_{m_+}^1(\bar w^2)\, \times \nonumber \\
& & \int_0^\infty\, d|\bar v|\,|\bar v|^3\,\exp(-\bar v^2)\,L_{k-1}^1(\bar v^2)
L_{m_+}^1(\bar v^2)\, - \, (m_+\longleftrightarrow m_+-1)\bigg]
 \nonumber \\
& = & is_f\,k\,\frac{B_f^4}{1024\,\pi^3}\;\delta_{kk'}\,(\delta_{m_+k-1} - \delta_{m_+k})\
,
\label{bk1}
\end{eqnarray}
where we have made use of the relation in Eq.~(\ref{master3}). Finally,
noting that
\begin{equation}
\sum_{m=0}^\infty\, s_f\,(\delta_{m_+k-1} - \delta_{m_+k})\,
g_{m,q_\parallel}^{+,f}\ = \ g_{k,q_\parallel}^{-,f} -
g_{k,q_\parallel}^{+,f}\ ,
\end{equation}
it is easy to see that Eqs.~(\ref{b1}), (\ref{bk0}) and (\ref{bk1}) lead to
our result in Eq.~(\ref{f2}).

\end{document}